\documentclass[aps,reprint,superscriptaddress,amsmath]{revtex4-1}

\usepackage{upgreek,graphicx}

\begin{document}

\title{Exploiting breakdown of the similarity relation for diffuse light transport: \\ simultaneous retrieval of scattering anisotropy and diffusion constant}

\author{Tomas Svensson}
\email[Email: ]{svensson@lens.unifi.it}
\affiliation{European Laboratory for Non-linear Spectroscopy (LENS), University of Florence, Via Nello Carrara 1, 50019 Sesto Fiorentino (FI), Italy}

\author{Romolo Savo}
\affiliation{European Laboratory for Non-linear Spectroscopy (LENS), University of Florence, Via Nello Carrara 1, 50019 Sesto Fiorentino (FI), Italy}

\author{Erik Alerstam}
\affiliation{Jet Propulsion Laboratory, California Institute of Technology, Pasadena, California 91109, USA}

\author{Kevin Vynck}
\affiliation{European Laboratory for Non-linear Spectroscopy (LENS), University of Florence, Via Nello Carrara 1, 50019 Sesto Fiorentino (FI), Italy}
\affiliation{Institut Langevin, ESPCI ParisTech, 1 rue Jussieu, 75005 Paris, France}

\author{Matteo Burresi}
\affiliation{European Laboratory for Non-linear Spectroscopy (LENS), University of Florence, Via Nello Carrara 1, 50019 Sesto Fiorentino (FI), Italy}
\affiliation{Istituto Nazionale di Ottica (CNR-INO), Largo Fermi 6, 50125 Firenze (FI), Italy}

\author{Diederik S. Wiersma}
\affiliation{European Laboratory for Non-linear Spectroscopy (LENS), University of Florence, Via Nello Carrara 1, 50019 Sesto Fiorentino (FI), Italy}
\affiliation{Istituto Nazionale di Ottica (CNR-INO), Largo Fermi 6, 50125 Firenze (FI), Italy}

\date{\today}

\begin{abstract}
As manifested in the similarity relation of diffuse light transport, it is difficult to assess single scattering characteristics from multiply scattered light. We take advantage of the limited validity of the diffusion approximation of light transport and demonstrate, experimentally and numerically, that even deep into the multiple scattering regime, time-resolved detection of transmitted light allows simultaneous assessment of both single scattering anisotropy and scattering mean free path, and therefore also macroscopic parameters like the diffusion constant and the transport mean free path. This is achieved via careful assessment of early light and matching against Monte Carlo simulations of radiative transfer.
\end{abstract}

\maketitle

Media having appearance and optical properties to a large extent determined by multiple scattering of light are truly abundant. This is reflected by an enormous multidisciplinary interest in the theory of light propagation in turbid media and experimental methods for accurate assessment of optical properties, structure and chemical composition of such systems. Applications range from astronomy \cite{Hansen1974_SpaceSciRev}, atmospherical science \cite{Mishchenko2006_Book} and climate research \cite{Davis2010_RepProgPhys} to medicine \cite{Welch2010_Book}, materials science \cite{Berne2000_Book} and fundamental photonics \cite{Akkermans2007_Book}. Despite being a field with a very long history, the complexity of radiative transfer has rendered approximative treatment prevalent, and there are still strong developments regarding accurate theoretical descriptions \cite{Liemert2011_PRA,Vitkin2011_NatCommun}. Generally, the similarity relation for diffusive light transport \cite{Hulst1980_Book,Graaff1993_OptEng} tells us that it is difficult to infer single scattering anisotropy ($g$) and scattering mean free path ($\ell_s$) from multiply scattered light \--- macroscopic transport is essentially governed by the transport mean free path, $\ell_t=\ell_s/(1-g)$. Measurement of microscopic properties are therefore normally made by studying light transport close to a source or transport through small samples (see, e.g.,\cite{Menon2005_PRL,Vitkin2011_NatCommun} and references therein).

Here, we take direct advantage of the limited validity of the diffusion approximation of light transport and show, numerically and experimentally, that time-resolved experiments allow assessment of single scattering characteristics even when signals are deep into the multiple scattering regime. Despite that time-resolved diffuse spectroscopy is extensively used to assess optical properties, structure and chemical composition of turbid media \cite{Chance1988_PNAS,Delpy1988_PhysMedBiol,Patterson1989_ApplOpt,Cubeddu1999_ApplPhysLett,Grosenick1999_ApplOpt,Schmidt2000_RevSciInstr,Ntziachristos2001_MedPhys,Johansson2002_ApplSpectrosc,Svensson2007_JBO,Pifferi2008_PRL,Svensson2011_PRL,Alerstam2012_PRE,DAndrea2012_ApplOpt}, as well as for fundamental investigations of light transport \cite{Watson1987_PRL,Kop1997_PRL,Wiersma2000_PRE,Sapienza2007_PhysRevLett,vanderBeek2012_PRB,Sperling2012_arXiv}, these opportunities has remained unexplored. Normally, the technique is considered to operating in the diffusion regime, and a major theoretical and experimental issue has instead been to understand to what extent diffusion theory is applicable \cite{Yoo1990_PRL,Yoo1990_OptLett,Kim1998_ApplOpt,Elaloufi2004_JOSAA,Alerstam2008_JBO,Alerstam2008_OptExpress}. Even in cases where diffusion breaks down and refined modeling has been used, evaluation has been limited to parameters accessible in the diffusion regime \cite{Pifferi1998_ApplOpt,Alerstam2008_JBO,Alerstam2008_OptExpress,Svensson2008_JBiophoton,Bouchard2010_OptExpress,Garofalakis2004_JOptA}. Since single scattering properties often are of great diagnostic value in optical characterization, an extension of capabilities to include also these  is important.

In radiative transfer, transport is modeled using an absorption coefficient $\mu_a$, a scattering coefficient $\mu_s$ and a scattering phase function. The famous transport equation follows from energy conservation \cite{Mishchenko2006_Book,Welch2010_Book}. The derivation is phenomenological, reducing the complex wave propagation into a Poissonian random walk with exponentially distributed steps between scattering events, and with a correlation in directionality given by the phase function (cf. \cite{Mishchenko2006_Book} for a treatment of the relation between Maxwell's equations and radiative transfer). The average step between scattering events is the scattering mean free path, $\ell_s=1/\mu_s$. On a macroscopic scale (i.e., after many scattering events), the transport equation reduces to a diffusion equation. Assuming homogenous random distribution and orientation of scatterers, the diffusion constant is
\begin{align}\label{eq:D}
   D = \frac{1}{3}\times\frac{\ell_s}{1-g}\times v_E 
\end{align}
where $g$ is the average cosine of the scattering angle, and $v_E$ the energy velocity \cite{Lagendijk1996_PhysRep,Pierrat2006_JOSAA}. Here we have reached the similarity relation mentioned earlier \--- as long as the quantity $\ell_t=\ell_s/(1-g)$ is conserved, macroscopic transport is similar. It is thus possible to map a diffusive process onto a random walk of exponentially distributed, isotropic and independent steps of average length $\ell_t$. The similarity has also been exploited to, e.g., relax the knowledge required on $g$ (and phase function) when using Monte Carlo simulations (MC) to evaluate time-resolved diffuse spectroscopy \cite{Kienle1996_PhysMedBiol,Pifferi1998_ApplOpt,Alerstam2008_JBO,Alerstam2008_OptExpress,Svensson2008_JBiophoton,Bouchard2010_OptExpress}. Having this said, the possibility to resolve the ambiguity related to the similarity relation by looking at early time-of-flights (TOF) was discussed already in the classic paper of Patterson, Chance and Wilson \cite{Patterson1989_ApplOpt}. In addition, there are several contributions on how  $g$ impacts the validity of the diffusion approximation \cite{Graaff1993_OptEng,Kienle1996_PhysMedBiol,Durian1997_JOSAA,Kim1998_ApplOpt,Garofalakis2004_JOptA,Alerstam2011_PhDThesis}, illustrating the intimate link between break in similarity and diffusion breakdown. Interestingly, it is widely assumed that diffusion theory is most accurate for $g=0$, while recent elaboration shows that the fastest convergence to a Gaussian spatial distribution, i.e. diffusion, occurs for $g\approx 0.7$ for the particular case of the Henyey-Greenstein phase function \cite{Alerstam2011_PhDThesis}. In any case, the breakdown of diffusion theory has most often been looked upon as a complication rather than an opportunity. In fact, it often recommended to exclude early times during analysis, partly due to the breakdown of diffusion theory, partly due to experimental uncertainties in timing \cite{Ntziachristos2001_MedPhys}. While this may be wise under certain circumstances, our work aims at demonstrating that it is possible to go beyond that perspective.

\begin{figure}[h]
  \centering
  \includegraphics[height=38mm]{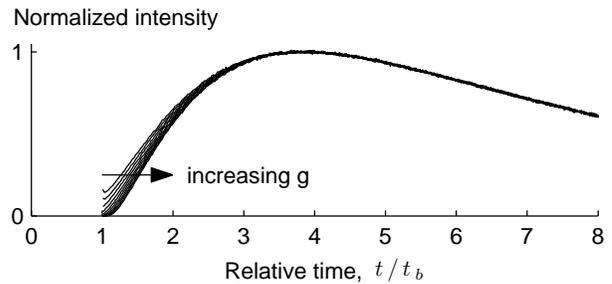}\\
  \caption{\small Simulated time-resolved total transmission for a turbid slab with $L/\ell_t=8$ for $g=0.0,0.1,...,0.9$ ($n_i$=1.5, $n_e$=1). Clearly, even deep into the multiple scattering regime, early dynamics can exhibit a significant dependence on $g$. This similarity breaking is clear evidence of diffusion breakdown. That a larger $g$ (more forward scattering) reduce that amount of early light is related to that the unscattered intensity decays with depth $z$ as $\exp(-z/\ell_s)$, and that having $g$ approach one while keeping $\ell_t$ fixed is accompanied by a reduction of $\ell_s$.}\label{FIG_SlabExample}
\end{figure}

We start with a theoretical view on the possibility to infer $g$ from TOF distributions. Although analytical solutions to the radiative transfer equation are becoming available \cite{Dominguez2010_ApplOpt,Liemert2010_OptLett,Liemert2011_PRA}, we will rely on Monte Carlo simulations (MC). MC is a straight-forward and well established method to solve the transport equation, capable of handling arbitrary boundaries, phase functions and source terms. When adhering to the new trend of using graphical processing units (GPUs) to parallelize computations \cite{Alerstam2008b_JBO}, simulations are done rapidly. A simulation involving injection of $10^8$ random walkers injected into a slab of thickness $L=8\ell_t$, as done here, takes on the order of one second on a modern GPU. Our  simulations were conducted on a GPU using the code in \cite{Alerstam2008b_JBO}, but modified for a slab geometry. We set absorption to zero and, as customary, used the Henyey-Greenstein phase function. We start by considering the case of an internal refractive index of $n_i$=1.5 and an external of $n_e$=1. The influence of $g$ on TOF distributions for total transmission is shown in Fig. \ref{FIG_SlabExample}. The dependence on $g$ is evident, but it is also clear that dynamics fall under similarity at longer times. Systems with strong forward scattering ($g\geq 0.7$) exhibit similar dynamics also for early times, meaning that the similarity relation breaks down when moving towards isotropic scattering (this in agreement with earlier works, cf. \cite{Kienle1996_PhysMedBiol,Alerstam2011_PhDThesis}). In this case $L/\ell_t=8$, so diffusion theory can be expected to rather well describe the long-time behavior \cite{Elaloufi2004_JOSAA}.  The time-integrated transmission was very close to 27.0\% for all values of $g$ , in good agreement with the prediction of diffusion theory \cite{Contini1997_ApplOpt,Vellekoop2005_PRE},  $T_D=(\ell_t+z_e)/(L+2z_e)=26.6$\%. Similarly, in terms of the ballistic time $t_b$, the decay time constant $\tau$ varied from $5.70\,t_b$ ($g=0.9$) to $5.78\,t_b$ ($g=0.0$), which is in fair agreement with diffusion theory, $\tau_D=(L+2z_e)^2/(\pi^2D)=6.26\,t_b$. 

\begin{figure}[h]
  \centering
  \includegraphics[height=40mm]{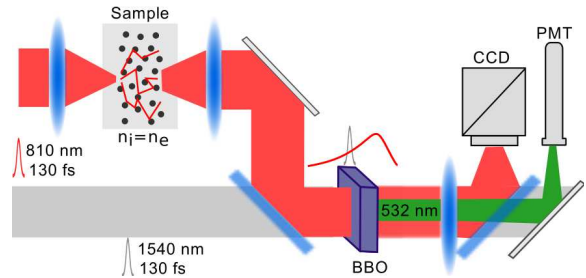}\\
  \caption{\small Ultrashort pulses at $\lambda=810~$nm from a titanium-sapphire laser are injected in a small spot on the sample, and the central part ($30~\upmu$m spot, monitored by CCD) of the transmission is collected and resolved in time using an ultrafast gating technique. Femtosecond time-resolution is achieved by overlapping the transmitted pulse with a fs reference pulse on a BBO crystal and monitoring sum-frequency generation at 532 nm (a variable delay line gives access to the full TOF distribution).}\label{FIG_Setup}
\end{figure}

Let us now turn to our experimental investigation. We have studied the time-resolved transmission of light ($\lambda=810~$nm) through a $252~\upmu$m thick turbid slab consisting of 1 vol\% nano-particles embedded in a polymer with $n$=1.5. The slab is located between glass slides, meaning that we are working with index-matched boundary conditions ($n_e$=$n_i$=1.5). The used scatterers (coated titania-particles, Huntsman Tioxide R-XL, USA) have a nominal diameter of $\sim$280 nm and an average refractive index of about 2.4. Mie calculations show that the single scattering anisotropy $g$ is around 0.6. The optical setup used for investigations of this sample is illustrated and described in Fig. \ref{FIG_Setup}. 
Note that, contrary to the total transmission configuration considered above, only central light is collected here. The approach described in this work is, in fact, very general and applies well to different configurations (this is further elaborated below).

\begin{figure}[h]
  \centering
  \includegraphics[width=80mm]{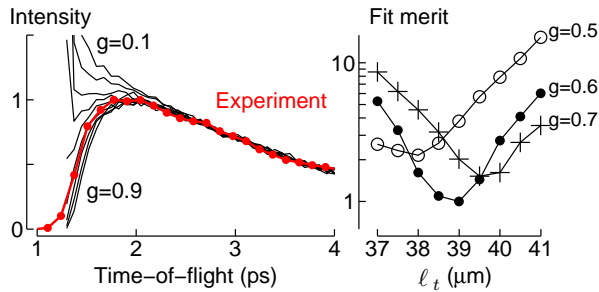}\\
  \caption{\small Evaluation of experimental data based on MC. The left graph shows simulations for $\ell_t=39~\upmu$m and different levels of $g$ (black lines, $g=0.1,\ldots,0.9$) and compares it with the experimental data. Experimental data is normalized to overlap with MC at late times where difference due to $g$ are negligible. The right graph shows obtained fit merits for the three best-matching $g$, revealing that the best least-squares fit is obtained for $g=0.6$ and $\ell_t=39~\upmu$m. The fitted $\ell_t$ is in good agreement with measurement on thicker slabs of the same material, and the fitted $g$ is in good agreement with Mie calculations. }\label{FIG_Experiment}
\end{figure}

Evaluation of obtained experimental data against a database of MC simulations for different levels of $\ell_t$ and $g$ (using only centrally transmitted walkers, in order to match experiment), with resolution $0.5~\upmu$m and 0.1 respectively, reveals an optimal match for $g=0.6$ and $\ell_t=39~\upmu$m (i.e., $\ell_s=15.6~\upmu$m). Fig. \ref{FIG_Experiment} shows the experimentally recored TOF distribution along with simulations of different levels of $g$ for $\ell_t=39~\upmu$m, as well as the result of least squares matching. 
Fit merits indicate that the precision is better than $\pm 0.1$ for $g$, and $\pm 1~\upmu$m for $\ell_t$. The marked difference in $g$-dependence between Fig. \ref{FIG_SlabExample} and Fig. \ref{FIG_Experiment} is related to that the former considers total transmission, while the latter concerns detection of light from a small area around the optical axis. Throwing away a large part of the diffuse transmission naturally result in a different balance between the ballistic and quasi-ballistic component and the multiply scattered part. Note also that $L/\ell_t$ is about 6.5, so established rules of thumb \cite{Elaloufi2004_JOSAA} implies that diffusion theory should fail to describe even late time behavior. However, the late-time decay of the total transmission obtained from MC ($\tau$=$3.6\pm0.1~$ps) is in good agreement with the predictions of diffusion theory ($\tau_D$=$3.60~$ps). The reason behind this agreement is that the validity of diffusion theory decreases with the refractive index contrast \cite{Elaloufi2004_JOSAA}, and we are now studying the forgiving case $n_i=n_e$.

In conclusion, we have shown that single scattering characteristics such as $g$ and $\ell_s$ are accessible also after long propagation distances. This is potentially a useful development, and we hope that our findings will stimulate further investigations of how to proceed beyond the similarity relation. The accuracy and precision of the approach depends on the transport regime (e.g. $L/\ell_t$) and the details of angular collection and conversion (cf. \cite{Yoo1990_OptLett}), being an important matter to be addressed in future research. Finally, it should be noted that measurements do not require ultrafast setups \textit{per se}, as required temporal stability and resolution depends on the scale of the system. We therefore anticipate widespread applicability of the approach, from characterization of turbid liquids and porous media to biological tissues.

\end{document}